\documentclass[11pt]{article}
\usepackage{moriond,epsfig}

\def\Journal#1#2#3#4{{#1} {\bf #2}, #3 (#4)}

\def\PRL{\em Phys. Rev. Lett.}
\def\PRD{{\em Phys. Rev.} D}

\def\be{\begin{equation}}
\def\ee{\end{equation}}
\def\bea{\begin{eqnarray}}
\def\eea{\end{eqnarray}}

\begin{document}
\vspace*{4cm}
\title{BEYOND EINSTEIN: COSMOLOGICAL TESTS OF MODEL INDEPENDENT MODIFIED GRAVITY}

\author{D.B. THOMAS }

\address{Theoretical Physics, Blackett Laboratory, Imperial College, London, England SW7 2AZ}

\maketitle\abstracts{Model-independent parametrisations of modified gravity have attracted a
lot of attention over the past few years; numerous combinations of experiments and observables
have been suggested to constrain these parameterisations, and future surveys look very promising.
Galaxy Clusters have been mentioned, but not looked at as extensively in the literature as some
other probes. Here we look at adding Galaxy Clusters into the mix of observables and examine
whether they could improve the constraints on the modified gravity parameters. In particular, we
forecast the constraints from combining the Planck CMB spectrum and SZ cluster catalogue and a
DES-like Weak Lensing survey. We've found that adding cluster counts improves the constraints obtained from combining CMB
and WL data. }

\section{Introduction }
Einstein's General Relativity (GR) is one of the principal ingredients
of modern cosmology. Nonetheless, it is our job as physicists to continue to test
even the fundamental pillars of cosmology in order to refine, improve
and further justify our model of the universe. Testing GR outside of the solar system can be quite challenging,
particularly as the effects of a different theory of gravity could be
degenerate with different possible constituents of the universe. This
is the case with the current cosmological observations that suggest the presence of
dark matter and dark energy. As well as explaining these observations, there are also
fundamental physics reasons for considering different theories of
gravity: GR is inconsistent with quantum mechanics and the search for
"Quantum Gravity" is one of the holy grails of modern physics. \\
Here, we are interested in testing deviations from GR in a model
independent way. There are several advantages to a model independent
approach; some alternatives to GR do exist but there is no complete
theory of, for example, quantum gravity to draw on. Also, amongst the many options there are no
"stand-out" candidates that are universally considered to be strong
alternatives. If a model independent approach suggests that the data is inconsistent with
concordance cosmology and GR, it will be relatively unambiguous and therefore a
strong motivator to develop alternative theories, as well as possibly
giving us a clue as to the nature of these theories.\\
There are studies in the literature on the constraining power of current data \cite{gbz10,bean10}
and the general conclusion is that the concordance cosomology
is consistent with all of the current data, although the data isn't
strongly constraining. Work has also gone into
forecasting future
experiments \cite{serra09,gbz09} and again there is a fair degree of consensus here, namely
that future surveys will greatly improve prospects. In this work we will examine the
constraints that can be put on model independent modified gravity using the combination of Cosmic Microwave Background anisotropies (CMB)
cross-correlated with weak lensing surveys and galaxy cluster counts. For more details and up to date figures, see \cite{me}.

\section{Modified Gravity}
Our potentials are defined in a flat FRW metric in the Newtonian gauge by $g_{00}=-1-2\Psi(\vec x , t)\,,g_{0i}=0\,,g_{ij}=a^2\delta_{ij}(1-2\Phi(\vec x ,t))\,.$
$\Psi$ is the Newtonian potential and is responsible for the
acceleration of massive particles. $\Phi$ is the curvature potential,
 which contributes to the acceleration of relativistic particles only.\\
Several sets of modified gravity parameters (MGPs) have been proposed, see \cite{huig} for one of the
first papers and \cite{mgps} for a partial translation table between the different parameterisations In this work we will use two MGPs, $\eta$ and $\mu$, following \cite{mgcamb} and implemented in the code MGCAMB, to describe departures from GR. The first, $\eta$, is the ratio of the two metric potentials,
$\eta=\Psi/\Phi$. This will be approximately unity in GR unless any of
the particle species has large anisotropic stress. The second, $\mu$, is a modification of
the poisson equation, and is essentially a time and space dependent
Newton's constant. Fourier expanding the spatial dependence with
wavenumbers $k$ and assuming isotropy, the modification of the
Possion equation is as follows
\begin{equation}\label{eq:mgp}
k^2 \Psi(a,k)=-4\pi G a^2 \mu(a,k) \rho (a)\Delta(a,k) \,,
\end{equation}
where, $a$ is the FRW scale factor, $G$ is Newton's constant, $\rho$
is the background density of cold dark matter and $\Delta$ is the
gauge invariant density contrast.\\
We will assume that
GR is valid up to a specified redshift $z_{mg}=30$. Beyond this, we assume that the MGPs transition to a constant value that
is different to the GR value. The background expansion history is already constrained to be close to
that of a $\Lambda$CDM model, we will therefore assume that the
modified gravity mimics the expansion history of a standard
$\Lambda$CDM setup.

\section{Observables}
\subsection{Cluster Counts}
Galaxy clusters are some of the largest collapsed structures in the
universe. According to the standard $\Lambda$CDM cosmology, they
typically consist of hot gas bound in a large cold dark matter
halo. They are a useful cosmological probe as their
size corresponds to scales near the linear to non--linear transition
in the underlying dark matter power spectrum. This has several
consequences: they probe the tail of the matter perturbation spectrum
and are therefore a sensitive probe of growth. In addition, galaxy
cluster counts can be predicted accurately from the linear theory matter power spectrum, using
semi-analytic formulae or ones calibrated from N-body
simulations.\\
Our theoretical predictions for the number of clusters in redshift
bins will be compared to predicted SZ catalogues for a number of
future observational stages. The SZ effect \cite{sz1} is a nearly
redshift independent tracer of clusters that is due to the
rescattering of CMB photons by hot intracluster gas. The observational
limits on SZ observations are, in principle, determined simply by
resolution and sky coverage.

\subsection{CMB}
With the release of Planck satellite \cite{planck} results only a few years away we
are entering an era where observations of the CMB total intensiy
spectrum will have reached the sample variance limit throughout scales
where primary effects dominate the signal. The sensitivity to MGPs in the CMB spectrum is restricted to the
largest scales and the main signal that will
arise on these scales is the ISW effect. This is sourced as
the Universe transitions into a dark energy dominated model and the
potential starts to decay. The effect can be described by the integral
of the time-deritvative of the sum of metric potentials along the line
of sight.

\subsection{Weak lensing}
The third observable we will use is the convergence power spectrum
from weak lensing surveys. Weak lensing is a relatively new
cosmological tool and is a measure of the small distortions of
background galaxies caused by gravitational lensing by large scale
structure \cite{kaiser92,refregier03}. Distortions of
individual bckground galaxies are virtually impossible to measure due
to the intrinsic ellipticity of galaxies. However, statistical results
averaging over large numbers of galaxies are now routinely reported.\\
For our initial weak lensing survey, we consider a DES-like
survey. DES \cite{des} (Dark Energy Survey) is a ground based survey that is scheduled to begin
observations in 2011. It will survey 5000 sq deg over 5 years and aims
to constrain dark energy with 4 probes: supernovae, BAO, galaxy
clusters and weak lensing, the latter being the probe we are
interested in here.
 
\section{Forecasts}
In this Section we carry out forecasts for two future
observational `Stages'. For weak
lensing and cluster counts we will assume two distinct observational
stages corresponding to short and long term development of
survey sizes and accuracies. This is unnecessary for the CMB as the data
from Planck over the range of interest will be cosmic variance limited and therefore essentially as
good as theoretically possible.\\
Stage 1 corresponds to a Planck-like SZ survey
and a DES-like weak lensing survey. DES will be
carried out on the Cerro Tololo Inter-American Observatory in the
Chilean Andes and should start taking data in late 2011. The stage 2 weak lensing survey is based on the
LSST \cite{lsst}, due to begin taking data in 2020. The stage 2 SZ survey corresponds to a
Planck-like survey, but with a better flux resolution, allowing smaller mass clusters to be detected.\\
Our forecasts will be based on Fisher matrix \cite{fisher1,fisher2} estimates of errors in a
subset of parameters comprising the MGPs $\eta$ and $\mu$ and two
parameters from the standard model that are expected to be most
correlated with them, namely, the total matter density $\Omega_m$ and
the primordial amplitude of scalar curvature perturbations $A$.

\section{Results}
\begin{figure}[ht]
\begin{minipage}[b]{0.5\linewidth}
\centering
\psfig{figure=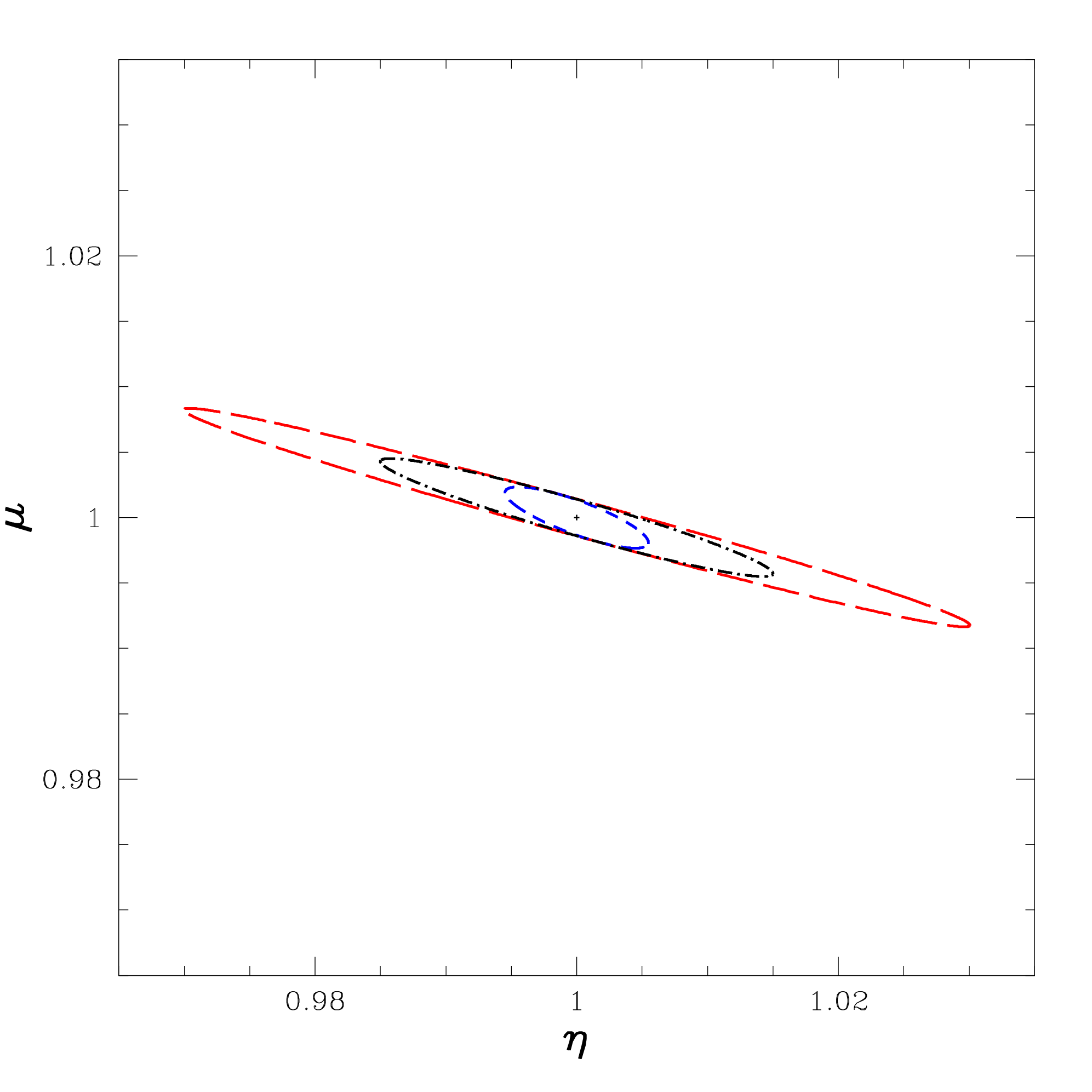,height=2in}
 \caption{Fisher constraints on $\eta$ and $\mu$ from combined CMB and
  weak lensing (including cross-correlations), red (dashed). The
  improvement obtained by adding cluster counts is seen in the blue (short-dashed)
  ellipse. When self--calibration uncertainties of the cluster data
  are included the constraints are weakened slightly (black,
  dash-dotted). All cases are for Stage I.}
 \label{fig:selfcal}
\end{minipage}
\hspace{0.5cm}
\begin{minipage}[b]{0.5\linewidth}
\centering
\psfig{figure=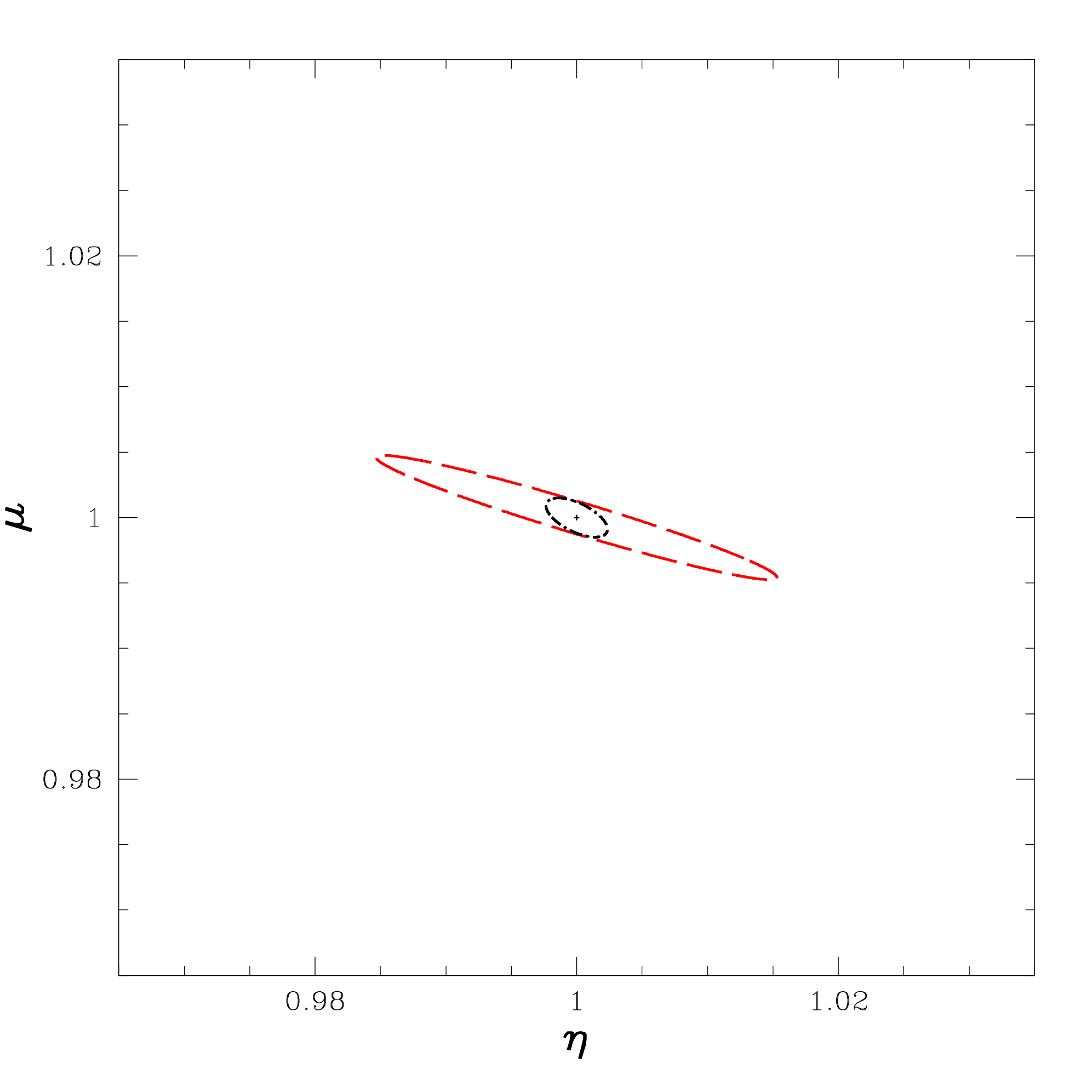,height=2in}
 \caption{Fisher constraints on $\eta$ and $\mu$ from combined CMB and
  weak lensing (including cross-correlations), red (dashed). The
  improvement obtained by adding cluster counts is seen in the black (dash-dotted)
  ellipse. All cases are for Stage III.\vspace{0.71cm}}
 \label{fig:jenk3}
\end{minipage}
\end{figure} 
\begin{table*}[ht]
\caption{1$\sigma$ constraints for first stage experiments}   
\centering                          
\begin{tabular}{|c c c c|}            
\hline                      
Parameter & CMB and WL cross-correlation & Clusters added & Cluster counts self calibrated\\ [0.5ex]   
\hline                              
$\Omega_m$ & 0.00104 & 0.00104 & 0.00104 \\   
Log ($10^{10}$ A) & 0.00215 & 0.00214 & 0.00214 \\
$\eta$ & 0.0300  & 0.00548 & 0.0150 \\
$\mu$  & 0.00835 & 0.00234 & 0.00451 \\[1ex]         
\hline                              
\end{tabular}
\label{table:jenk1}          
\end{table*}
After cross-correlating the CMB with weak lensing, the constraints on the MGPs are quite
good. This is due to the complementarity between the data sets; with the CMB providing strong
constraints on the standard parameters, any degeneracies between the standard parameters and the
MGPs in the weak lensing data are broken. However, since both the CMB and weak lensing rely on the sum of the two potentials,
there is still a degeneracy between the MGPs that is affecting the constraints. This is where the galaxy
cluster counts are useful, as only $\Psi$ is relevant and hence only $\mu$ contributes. Thus, the data from the
cluster counts breaks the degeneracy between the MGPs from the CMB and Weak lensing,
creating a much tighter constraint as shown in figure \ref{fig:selfcal}. There are some uncertainties associated with cluster counts \cite{limahu05}, and these are also shown in
figure \ref{fig:selfcal}. Although marginalising over these uncertainties reduces the impact of clusters, clusters still add to the constraining power of the CMB and Weak lensing. The constraints on
the parameters for the first stage of experiments are shown in table \ref{table:jenk1}. In addition, figure \ref{fig:jenk3} shows how galaxy clusters are still a worthwhile addition to
cross correlated CMB and weak lensing measurements for the longer term survey.

\section{Conclusion}
Over the next 5-10 years, deviations from GR
should be well constrained, and the concordance cosmology will either be more
secure or may even have undergone a paradigm shift. If the latter is the case, then the results from the model independent tests could be crucial
in helping to find a new theory of gravity.


\section*{References}
\begin{thebibliography}{99}

\bibitem{me}D.B.Thomas {\it et al}, \Journal{ArXiv e-prints}{astro-ph}{1107.0727}{2011}
\bibitem{gbz10}G.B. Zhao {\it et al}, \Journal{\PRD}{81}{103510}{2010}
\bibitem{bean10}R. Bean {\it et al}, \Journal{\PRD}{81}{083534}{2010}
\bibitem{serra09}P. Serra {\it et al}, \Journal{\PRD}{79}{101301}{2009}
\bibitem{gbz09}G.B. Zhao {\it et al}, \Journal{\PRL}{103}{241301}{2009}
\bibitem{huig}W Hu {\it et al}, \Journal{\PRD}{76}{104043}{2007}.
\bibitem{mgps}S.F. Daniel {\it et al}, \Journal{\PRD}{81}{123508}{2010}
\bibitem{mgcamb}G.B. Zhao {\it et al}, \Journal{\PRD}{79}{083513}{2009}.
\bibitem{sz1}R.A. Sunyaev {\it et al}, \Journal{Comments on Astrophysics and Space Physics}{2}{66}{1970}
\bibitem{planck}The Planck Collaboration, astro-ph/0604069
\bibitem{kaiser92}N. Kaiser, \Journal{Astrophys.J.}{388}{272}{1992}
\bibitem{refregier03}A. Refregier, \Journal{Ann.Rev.Astron.Astrophys.}{41}{645}{2003}
\bibitem{des}http://www.darkenergysurvey.org
\bibitem{lsst}http://www.lsst.org/lsst
\bibitem{fisher1}M. Tegmark,\Journal{\PRL}{79}{3806}{1997}
\bibitem{fisher2}M. Tegmark {\it et al},\Journal{Astrophys.J}{480}{22}{1997}
\bibitem{limahu05}M. Lima {\it et al}, \Journal{\PRD}{72}{043006}{2005}
\end{thebibliography}
\end{document}